# Twisted Bilayer Graphene Superlattices

Yanan Wang[1], Zhihua Su[1], Wei Wu[1,2], Shu Nie[3], Nan Xie[4], Huiqi Gong[4], Yang Guo[4], Joon Hwan Lee[5], Sirui Xing[1,2], Xiaoxiang Lu[1], Haiyan Wang[5], Xinghua Lu[4], Kevin McCarty[3], Shin-shem Pei[1,2], Francisco Robles-Hernandez[6], Viktor G. Hadjiev[7], Jiming Bao[1,*]

[1]Department of Electrical and Computer Engineering
University of Houston, Houston, TX 77204, USA

[2]Center for Advanced Materials
University of Houston, Houston, TX 77204, USA

[3]Sandia National Laboratories, Livermore, CA 94550, USA

[4]Institute of Physics, Chinese Academy of Sciences, Beijing 100190, China

[5]Department of Electrical and Computer Engineering
Texas A&M University, College Station, Texas 77843, USA

[6]College of Engineering Technology
University of Houston, Houston, TX 77204, USA

[7]Texas Center for Superconductivity and Department of Mechanical Engineering, University of Houston, Houston, TX 77204, USA

*To whom correspondence should be addressed: jbao@uh.edu.

# Abstract

Twisted bilayer graphene (tBLG) provides us with a large rotational freedom to explore new physics and novel device applications, but many of its basic properties remain unresolved. Here we report the synthesis and systematic Raman study of tBLG. Chemical vapor deposition was used to synthesize hexagon-shaped tBLG with a rotation angle that can be conveniently determined by relative edge misalignment. Superlattice structures are revealed by the observation of two distinctive Raman features: folded optical phonons and enhanced intensity of the 2D-band. Both signatures are strongly correlated with G-line resonance, rotation angle and laser excitation energy. The frequency of folded phonons decreases with the increase of the rotation angle due to increasing size of the reduced Brillouin zone (rBZ) and the zone folding of transverse optic (TO) phonons to the rBZ of superlattices. The anomalous enhancement of 2D-band intensity is ascribed to the constructive quantum interference between two Raman paths enabled by a near-degenerate Dirac cone. The fabrication and Raman identification of superlattices pave the way for further basic study and new applications of tBLG.

One-dimensional superlattices typically consist of periodic alternation of two or more chemically distinctive material layers[1]. Graphene, a single-atom layer with a hexagonal lattice has provided us an opportunity to create new two-dimensional superlattices by stacking one layer on the top of the other[2]. Such bilayer graphene superlattices (BLGS) have attracted considerable attention because of the tunability of interlayer coupling and band structure resulting from the freedom of relative rotation[2-11]. While band structures of BLGS, in particular, at several commensurate rotation angles have been calculated by many groups[4-7, 9, 11-14], a comprehensive experimental confirmation of BLGS is still missing. Recent observations of G-line Raman resonance in twisted bilayer graphene (tBLG) with small rotation angles (< 16$^{o}$) revealed the rotation-dependent electronic band structures[9, 12, 13, 15], but such resonances can also be explained by the continuum model[13], which neglects the underlying superlattice structure and treats the band structure of tBLG as a superposition of rotated Dirac cones. Other groups reported Raman observation of new phonon modes indicative of bilayer superlattice structures[16-18], however, neither the relationship among these phonons, nor their relationship with the electronic band structure and G-line resonance of bilayer graphene is clearly established yet.

In this work we report the observation of new Raman features and their strong, systematic correlation with the crystal and electronic band structure of tBLG. A comprehensive picture of electronic and vibrational properties of tBLG was obtained by utilizing our unique bilayer samples and by our extensive Raman study that makes use of six laser lines covering a wide spectrum from near-infrared to UV. The G-line resonances were observed for the first time in tBLG with rotation angles larger than 20$^{o}$ using the 364-nm laser line. A rotation-dependent new folded phonon was observed in all the samples. This new phonon has a lower frequency than the G-line, and it can only be observed when the laser is tuned with the G resonance. Finally, a nearly four-fold increase in the 2D-band Raman intensity is observed when the laser is tuned below the G-line resonance. These correlated Raman spectra of G-line, 2D-band, and folded phonons reveal the underlying superlattice structure of tBLG. As synthesis using chemical vapor deposition (CVD) and Raman characterization of single-layer graphene have played an enormous role in rapidly advancing graphene research[19-21], our Raman identification of BLGS and CVD synthesis of isolated tBLG has enabled future research beyond single-layer graphene.

Twisted bilayer graphene was grown on Cu foils by CVD at ambient pressure in a quartz tube furnace[22]. The conditions are similar to what we used for single-layer graphene[21, 23], except that a larger flow rate of $CH_4$ was used to facilitate the growth of bilayer graphene. Figures 1a–d show examples of bilayer graphene islands that consist of two graphene hexagons stacked on each other. It can be seen that two layers can have relative rotation angles from 0 to nearly 30 degrees. The lattice rotation is confirmed by investigations of transmission electron microscopy (TEM) and scanning tunneling microscopy (STM), by which electron diffraction and lattice Moiré patterns can be seen in SF. 1 and 2 of the supporting online material. The effect of lattice rotation on the electronic band structure is also revealed by the STM observation of Van Hove singularities (VHS, see SF. 3)[24]. One important advantage of such bilayer graphene structures compared to those previously reported is that the relative edge misorientation of two hexagons represents their actual lattice rotation. This conclusion is obtained by comparing low-energy electron diffraction (LEED) with low-energy electron microscopy (LEEM) from the same region of graphene[22]. In the case of two domains contained in one hexagon, as shown in Fig. 1e and SF. 3b, the lattice rotation can still be determined from the misorientation between two neighboring edges. For instance, the smaller hexagon in Fig. 1e contains two domains delineated approximately by the dashed red line: the right domain has an AB-stacking with the larger hexagon, while the left domain has a ~13º relative rotation. This observation is also confirmed by Raman mapping as shown in Fig. 1f. The relationship between edge misorientation and lattice rotation has facilitated the identification of tBLG, where rotation angles in this work are all based on the average edge misorientations of available edges between two hexagons without further confirmation of LEED. Similar stacked graphene hexagons were reported before, but their lattice rotation was not verified with direct lattice imaging techniques such as TEM, STM or LEED[25].

As expected, such tBLG as shown in Fig. 1e should exhibit a strong G-line resonance when excited by a 532-nm laser[13, 15]. This prediction is confirmed by Fig. 1g, which shows Raman spectra from the single-layer and double-layer regions marked by a blue dot and a white dot, respectively. The dependence of G-line intensity on five different laser wavelengths shown in figures 1h–i further confirms the G-line Raman resonance near 532 nm, in agreement with previous observations and calculations[13, 15]. The 2D-band also shows an expected behavior: its intensity drops as the laser increases to higher excitation energies above the G-line resonance[15].

Despite these seemingly familiar behaviors for G-line and 2D-band, careful inspection reveals new Raman features not reported before. First, a new Raman peak appears at 1489 $cm^{-1}$ from the bilayer region. As can be seen in Fig. 1g, this Raman line is weak in intensity compared to the G-line of bilayer graphene, but it is comparable with the G-line of single-layer graphene. Raman spectra at five laser wavelengths in Fig. 1h show that this new line shares the same resonant behavior as the G-line: it is too weak to be observed when there is no G-line resonance under 638- or 458-nm laser lines. This new peak also shows no dispersion: its frequency is independent of the laser excitation energy. In addition, the bilayer 2D-band Raman intensity becomes more than doubled compared to that of single-layer graphene at the long wavelength of 638 nm, for which there is no G resonance.

Such a nondispersive new Raman peak is ubiquitous, and always observed whenever there is a G-line resonance. SF. 4 shows their Raman spectra in three different bilayer graphene structures with rotation angles of ~9°, ~11°, and ~16°. As anticipated, the position of G-line resonance shifts accordingly. Again, the same strong correlation between new Raman peaks and G-line is observed. However, unlike the G-line, the frequency of the new Raman line is strongly dependent on the rotation angle: its frequency decreases when the rotation angle increases. As we discuss later, these are the signatures of folded phonons in bilayer graphene superlattices. These folded phonons are very different from those reported before in many ways[16-18]. They appear in different frequency ranges, and also exhibit a unique resonant behavior with respect to the G-line and excitation energy.

The observation of more than two-fold enhancement in 2D-band intensity is also a general phenomenon provided that laser excitation energy is tuned below the G resonance. This observation can be best illustrated in a tBLG with such a large rotation angle that none of the visible laser lines meets the energy for G resonance. Such a tBLG with a rotation angle of ~25.6° is shown in the inset of Fig. 2a. Raman spectra in the same figure at 638-nm laser excitation show that the 2D-band intensity increases by more than four times. Figure 2b shows two other examples where enhanced 2D-bands are observed at 532 nm. The relatively smaller 2D-band enhancement for the tBLG with 15.9º rotation angle is due to the fact that the 532-nm laser comes closer to its G-line resonance excitation energy. Figure 2c shows the bilayer to single-layer intensity ratios for G-line and 2D-band at the five visible laser lines. The ratios become smaller as the excitation energy increases, but the 2D-band always increases more in intensity than the G line. As a comparison, for bilayer graphene made from simple mechanical stacking as shown in SF. 5, the 2D-band intensity of bilayer graphene is approximately twice that observed in single-layer graphene.

According to previous calculations, G-line resonance should be observed even for tBLG with a large rotation angle if the excitation energy is tuned high enough[13, 15]. This prediction is verified when a 364-nm UV laser was used. Figure 2d shows a more than 30 times enhancement in G-line intensity when the same 25.6º sample was excited by the UV line. It should be noted that UV lasers were used in the observation of phonons in twisted bilayer graphene[17, 18], but this is the first time that G-line resonance is observed. Similar to the spectral transition in 2D Raman shown in Fig. 1i, the integrated intensity of 2D-band becomes weaker under UV excitation. But where is the folded phonon? We show that the strong Raman peak near D-line position is actually the folded phonon.

This assignment of the folded phonon is based on the following observations. First, the position of this peak is dependent on the rotation angle of tBLG. Figures 3a–c show a series of UV Raman spectra in six different samples including two in Fig. 2b. It can be seen that the peak positions of 2D-band remain almost the same, but a large shift in the positions of Raman peaks near the D-line is observed. According to the relationship between D-line and 2D-band, the Raman shift of the D-line is simply half the Raman shift of the 2D-band. Thus, the strong Raman near D-line cannot be the usual D-line as observed by visible laser lines. Second, as before, such strong Raman peaks near D-line are only observed when there is G-line resonance. This is supported by all six samples shown in Fig. 3a–c. Because the 15.9º sample exhibits no G-line enhancement, there is no observable peak near the usual D-line. The other five samples show different degrees of G enhancement from 5 to 20 times. Unlike Raman with visible laser lines, the UV Raman D-line is very weak: the relatively strong D-line in single-layer graphene shown in Fig. 2d is due to the defects created by UV exposure during the Raman[26]. The dependence of observed phonon frequencies on the rotation angle is plotted in Fig. 3d. A general trend can be observed: a larger rotation angle leads to a lower phonon frequency.

The above observations of folded phonon modes and enhanced 2D-band, as well as their relationships with rotation angle and excitation energy, have revealed new information on the band structure of tBLG. Previously, two different models were developed to explain the G-line resonance: rotated Dirac cones and band structure of commensurate bilayer graphene[12, 13, 15]. Our observation of G resonance and VHS can be understood using either model, but the theory of rotated Dirac cones cannot account for the observed new Raman features. Because rotated Dirac cone theory assumes two nearly independent graphene layers in tBLG, the intensity of 2D-band in tBLG is simply twice that of single-layer graphene, as is shown in the calculated curve in Ref. 15, when the laser energy is tuned below the G-line resonance. More importantly, such a continuum model ignores the rotation-dependent superlattice structure, which is necessary for the emergence of folded phonons. As we discuss below, the folded phonon modes and enhanced 2D-band Raman are clear manifestation of the superlattice nature of twisted bilayer graphene.

Folded phonons are frequently observed in 1-D superlattices[1, 27]. The frequency of folded phonons can be estimated by zone folding of the initial phonon dispersion curve into the rBZ of the superlattices. Conversely, the character of folded phonon is an important measure of the quality of superlattice structure[1]. Figure 4a marks the positions of observed phonons on the transverse optical (TO) phonon curve along the Γ-K direction. The phonon at the Γ-point produces the well-known G-line line, which has the highest frequency. As crystal momentum increases, the phonon frequency decreases. Important features of these phonons can be understood from this dispersion curve without involving rigorous numerical calculation. As observed folded phonons fall in the frequency range of the TO phonon branch in the first BZ, their frequencies can be obtained through zone folding of monolayer TO phonons to the rBZ, possibly along a direction different from Γ-K due to the 2-D nature of bilayer superlattices. Because the size of the rBZ in general becomes larger for tBLG with a larger rotation angle, so does the crystal momentum of the folded phonon. Based on the TO phonon dispersion in Fig. 4a, the frequency of folded phonons should become smaller for a superlattice with a larger rotation angle.

Following the same zone-folding principle used in estimating phonon frequency in 1-D superlattices[1, 27], we can obtain the positions of folded phonons in 2-D graphene superlattices. Figure 4b shows BZs of single-layer graphene and rBZ of the bilayer superlattice with a rotation angle of 13.2°. Also shown are two sets of six-fold reciprocal lattices A and B of the superlattice. It is the position of these reciprocal lattices that determine the frequency of folded phonons. This can be understood in the following two ways, using reciprocal lattice A as an example. When TO dispersion in the larger BZ of single-layer graphene is mapped into the reduced BZ of the superlattice, A will be mapped to the Γ-point of the BZ, and becomes Γ-point optical phonons that can be probed by Raman scattering. Another way to say this is that with the assistance of crystal momentum Γ-A, the phonon at A is able to be accessed by Raman. Based on the six-fold symmetry, phonons at six reciprocal lattices equivalent to lattice A will have the same frequency. Its frequency can be estimated as follows if we neglect the dispersion warping of TO dispersion along different directions. Because the length of Γ-A is ~ 0.4 × Γ-K, a frequency of ~ 1480 $cm^{-1}$ is obtained, which is very close to the observed frequency of ~1489 $cm^{-1}$.

The rotation-dependent lattice structure has a strong effect on the 2D-band Raman as well. This is because 2D-band is a result of the double-resonance Raman process: its characteristic directly reflects the underlying electronic band structure of the superlattice. Band structure of bilayer superlattices have been calculated by many groups, especially for tBLG with commensurate angles such as 13.2°[4-7, 9, 11-14]. Because of the symmetry, the band structure of BLGS is still characterized by the reduced BZ, as shown in Fig. 4b. Although band structures of tBLG vary from one rotation angle to the other, they share many similar features. The Dirac band is two-fold degenerate from the Dirac point Γ up to the M-point of the reduced BZ, where the Dirac band begins to split[4-7, 9, 11-14]. It is important to note that the Dirac band is not perfectly degenerate: there is a finite gap between two Dirac bands. The size of the gap depends on interlayer coupling: a larger interaction leads to a larger gap[11]. It was calculated that interlayer coupling is related to the size of the superlattice cell: larger cells result in a weaker interlayer coupling. In any case, the splitting between two bands is more than 10 times smaller than the gap in AB-stacking graphene. It is this nearly degenerate Dirac band that is responsible for the enhanced 2D Raman[11].

Following the same procedure described in Ref. 19 for Raman scattering in AB-stacking graphene, we can construct four similar double-resonance Raman paths for the 2D-band. Figure 5 shows the schematic of Dirac bands along the K-M direction and four Raman paths associated with two major optical absorptions[19]. Let's focus on two Raman paths in Fig. 5a and discuss the consequence of degenerate band on the Raman intensity. The two Raman paths A→B→D and A→C→D are degenerate because of the degeneracy of two bands at B and C. The total Raman intensity is proportional to the modular square of the total amplitude $|A\rightarrow B\rightarrow D + A\rightarrow C\rightarrow D|^2$, which is twice the sum of $|A\rightarrow B\rightarrow D|^2$ and $|A\rightarrow C\rightarrow D|^2$ assuming two paths have the same Raman amplitude. This explains why 2D-band intensity in BLGS is nearly twice the total intensity of two independent monolayer graphenes, or four times that of single-layer graphene. As can be seen, this enhancement is purely due to quantum interference between two paths.

It is important to point out that it is the finite interlayer coupling and finite gap between Dirac bands that makes it possible to observe such quantum interference between two Raman paths. For a large gap, as in AB-stacking bilayer graphene, peaks from such two Raman paths can be well resolved, and two Raman paths are distinguishable, so there is no such quantum interference. Nor does it happen if two bands are perfectly degenerate. This can be understood in the following way: if two bands are completely degenerate, then there will be no interlayer coupling, meaning that two graphene layers are independent from each other. The Raman intensity of any lines should then be a simple sum of Raman intensity from the individual layers, or twice the Raman intensity of a single layer as shown in SF. 5. On the other hand, if there is no interlayer coupling, two bands shown in Fig. 5 are simply the bands of individual single-layer graphene folded into the rBZ. The Raman path A→C→D will then be missing because interband transition from A to C is not allowed any more.

Experimentally, the gap between two Dirac bands is too small to be resolved by Raman spectroscopy, and the 2D-band lineshape in tBLG is almost indistinguishable from that of single-layer graphene. Such monolayer-like 2D-band has been observed in tBLG and turbostratic graphene[12, 13, 15]. On one hand, such observation validates our assumption of the quantum interference between two indistinguishable Raman paths. On the other hand, it has led many people to believe that this 2D-band originates from single-layer graphene on the basis of either rotated Dirac cones or degenerate Dirac bands. Despite such theoretical understandings, the enhanced 2D-band has been frequently observed in those previous Raman studies[12, 13, 15]. Again, such intensity enhancement can only be observed when the laser energy is tuned below the G resonance. When the laser energy increases above the energy for G-resonance, the Raman contribution from the inner loop will be lost, resulting in a significant drop in 2D-band Raman intensity[28].

According to the zone folding shown in Fig. 4b, two different phonon modes are expected from reciprocal lattices A and B, but a lower-frequency phonon at B has not been observed. This is the case for all of the samples we have measured. We believe the following reasons explain why only the phonon with the highest frequency is observed, although an exact mechanism is still under investigation. Let's take the 13.2º tBLG as an example. First, the crystal momentum Γ-A is the fundamental reciprocal lattice vector of the superlattice, while Γ-B represents a higher-order crystal momentum. Raman scattering is typically much stronger for processes involving the fundamental mode than the higher order modes. Second, the phonon Γ-A is closer in energy to the G-line compared with the folded phonon Γ-B. The Γ-A phonon is thus more likely to be resonantly excited when the resonance condition for the G-line is satisfied.

Mathematically, a perfect bilayer graphene superlattice requires that two sub-lattices of single-layer graphene be commensurate. Such a condition can only be satisfied by a small set of discrete angles, which include 13.2º and 27.8º. A slight deviation from these commensurate angles, especially from the large angle of 27.8º, will make a bilayer graphene lattice either non-periodic or a new superlattice with a much larger lattice constant. Accordingly, either there is no folded phonon, or a folded phonon with a very different frequency. However, such a scenario does not agree with our observation. As can be seen in Fig. 3d, the frequency of folded phonons decreases gradually without abrupt fluctuation near the large rotation angle of 26º. This observation leads us to conclude that a tBLG without an exact commensurate rotation angle can be approximated as a superlattice that is commensurate and has the smallest size of supercell.

For a small rotation angle, the length of the reciprocal lattice of the superlattice can be expressed as $q(\theta) = \frac{8\pi}{\sqrt{3}a}\sin(\frac{\theta}{2})$, where θ is the rotation angle, and *a* is the lattice constant of monolayer graphene. *q(θ)* becomes the reciprocal lattice of original monolayer graphene when θ approaches 60º. This equation represents the reciprocal lattice of the superlattice with the smallest possible supercell for a given rotation angle. For a small rotation angle of less than 10 degrees, it is easy to find such a commensurate supercell within a small range of rotation angles.

Because of this, the equation can be used to estimate the frequency of folded phonons based on the TO dispersion shown in Fig. 4a. In this sense, the continuum model can still be used to estimate the phonon frequency of the superlattice. This method of estimation breaks down for large rotation angles. In the limit of the largest rotation angle of 30°, *q(θ)* approaches Γ-K, implying that the folded phonon shows a frequency at the K-point of ~1220 $cm^{-1}$. Clearly, this is not the case for our observation. It is well known that there is no such superlattice at a rotation of 30°; furthermore, the superlattice with the closest rotation angle and the smallest supercell happens at the rotation of 27.8°[11]. The frequency of folded phonons at such a rotation angle can still be estimated from the method shown in Fig. 4b.

It was speculated by the continuum model that the interlayer coupling becomes negligible for a tBLG with a large rotation angle [3]. But our observations indicate that this is not the case: a G-line resonance and a relatively strong folded phonon are observed for large-angle twisted graphene. This observation agrees with the calculation based on a graphene superlattice, and it is a result of a smaller size of unit cell of the superlattice for larger rotation angles [11]. The fact that both G-line resonance and folded phonons are observed for any studied bilayer graphene islands implies that tBLG synthesized by CVD always takes the most stable near-perfect lattice configuration with the smallest lattice constant, which is supported by a continuous decrease of folded phonon frequency as the rotation angle increases.

In conclusion, we synthesized isolated bilayer graphene hexagons and identified their superlattice structures using TEM, STM, LEED, LEEM, and Raman scattering. We obtained comprehensive and unique Raman signatures of superlattices, and correlated them with the electronic band structure of bilayer superlattices. Such Raman characterization provides a solid basis for further understanding and exploring rich physics and other unique properties of bilayer graphene. Single-layer graphene, a one-atom-thick two-dimensional lattice, thus provides us a new building block to engineer new two-dimensional structures with tunable properties.

Acknowledgements:

The work at Sandia National Laboratories was supported by the Office of Basic Energy Sciences, Division of Materials Sciences and Engineering of the US DOE under contract No. DE-AC04-94AL85000. SSP, JMB, WW and SRX acknowledge support from the Delta Electronics Foundation and UH CAM. JMB acknowledges support from the Robert A Welch Foundation (E-1728) and the National Science Foundation (DMR-0907336, ECCS-1240510).

**Figure captions:**

Figure 1. Characterizations of rotationally twisted bilayer graphene (tBLG) by scanning electron microscopy (SEM) and Raman scattering. (a–d) SEM pictures of representative tBLG hexagons with various rotation angles. Scale bars: 5 µm. (e–i) SEM and Raman characterizations of a ~13º tBLG. (e): SEM image. The blue and white dots indicate the regions where Raman was measured for monolayer and bilayer graphene. The red dashed line delineates the ~13º tBLG from AB-stacking bilayer domain. (f): Raman mapping with a 532-nm laser line. (g): Raman spectra of monolayer (1L, black curve) and bilayer (2L, red curve) graphene illustrating the G-line resonance under 532-nm laser excitation. (h) Dependence of a new Raman peak (folded phonon) on laser excitation energy. The spectra are normalized relative to the G-line of single-layer graphene. (i) Raman intensity ratios of bilayer to single-layer for G-line (blue) and 2D-band (red) as a function of laser excitation energy.

Figure 2. Raman spectra of tBLG with large rotation angles. (a): Raman spectra from single-layer (1L) and bilayer (2L) regions of a ~25.6º tBLG at 638-nm laser excitation. Inset: SEM image. (b): Raman spectra from monolayer graphene (1L, black) and bilayer graphene regions in a ~26.6º (red) and a ~15.9º (blue) tBLG under 532-nm laser excitation. Inset: corresponding SEM images. (c): Bilayer to single-layer Raman intensity ratios for G-line (black) and 2D-band (red) of the ~25.6º tBLG under different laser excitation energies. (d) Raman spectra from single-layer (black) and bilayer (red) regions of the ~25.6º tBLG under 364-nm laser excitation.

Figure 3. (a–c): 364-nm UV Raman spectra of the ~15.9º (black) tBLG and five other tBLGs with rotation angles greater than 21°. (a): Raman spectra indicating G-line resonance with an enhancement of 5 to 20 times for large-angle tBLGs. There is no G-line resonance for the ~15.9º tBLG. (b): Close-up view of (a) near 2D-band. (c): Close-up view of (a) near 1400 $cm^{-1}$. (d): Observed phonon frequency as a function of rotation angle.

Figure 4. Estimation of folded phonon frequency from TO dispersion of single-layer graphene. (a) Relationship between the observed frequencies of folded phonons and TO dispersion curve along Γ-K direction. (b) Reduced Brillouin zone (rBZ) and reciprocal lattices of 13.2º twisted bilayer graphene in the BZ of single-layer graphene. The phonon marked by A can be excited by Raman with the exchange of crystal momentum Γ-A.

.

Figure 5. Schematic of 2D-band double-resonance Raman paths enabled by the band structure of a bilayer graphene superlattice. Only Raman contributions from inner loop and transitions involving electrons are shown. Two degenerate Raman paths associated with the optical transitions between two inner branches (a) and two outer branches (b) of Dirac bands.

Fig. 1.

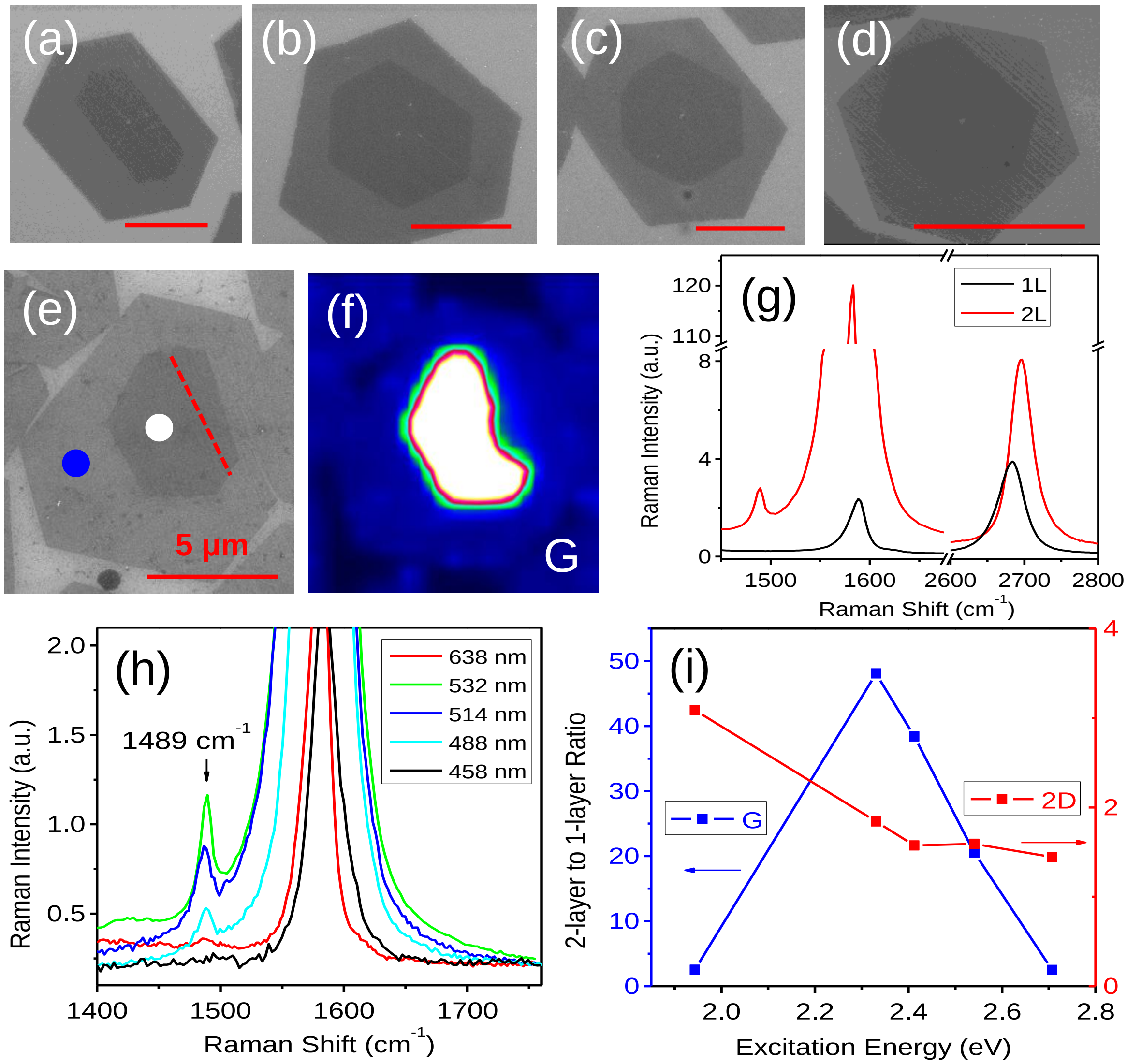

Fig. 2.

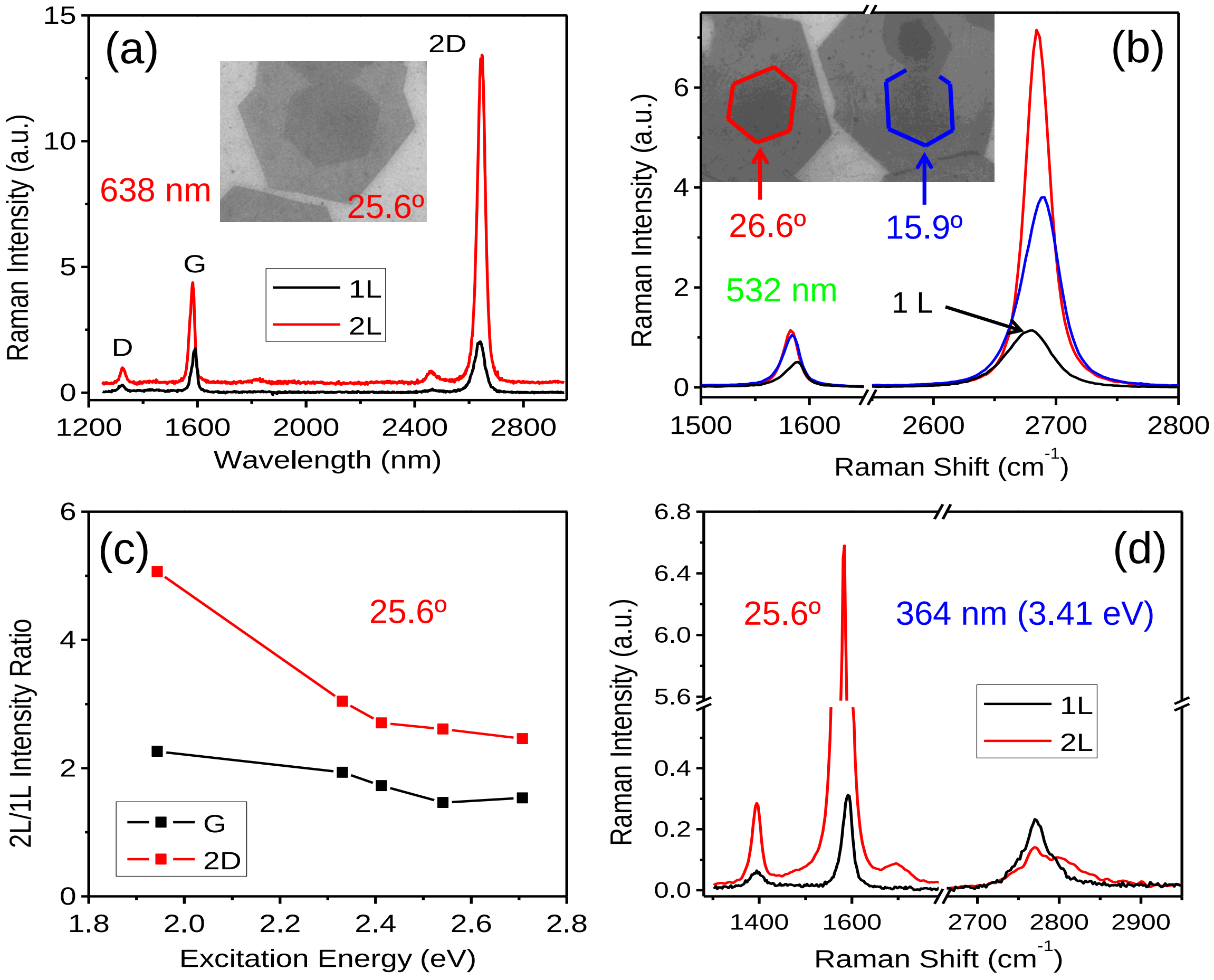

Fig. 3.

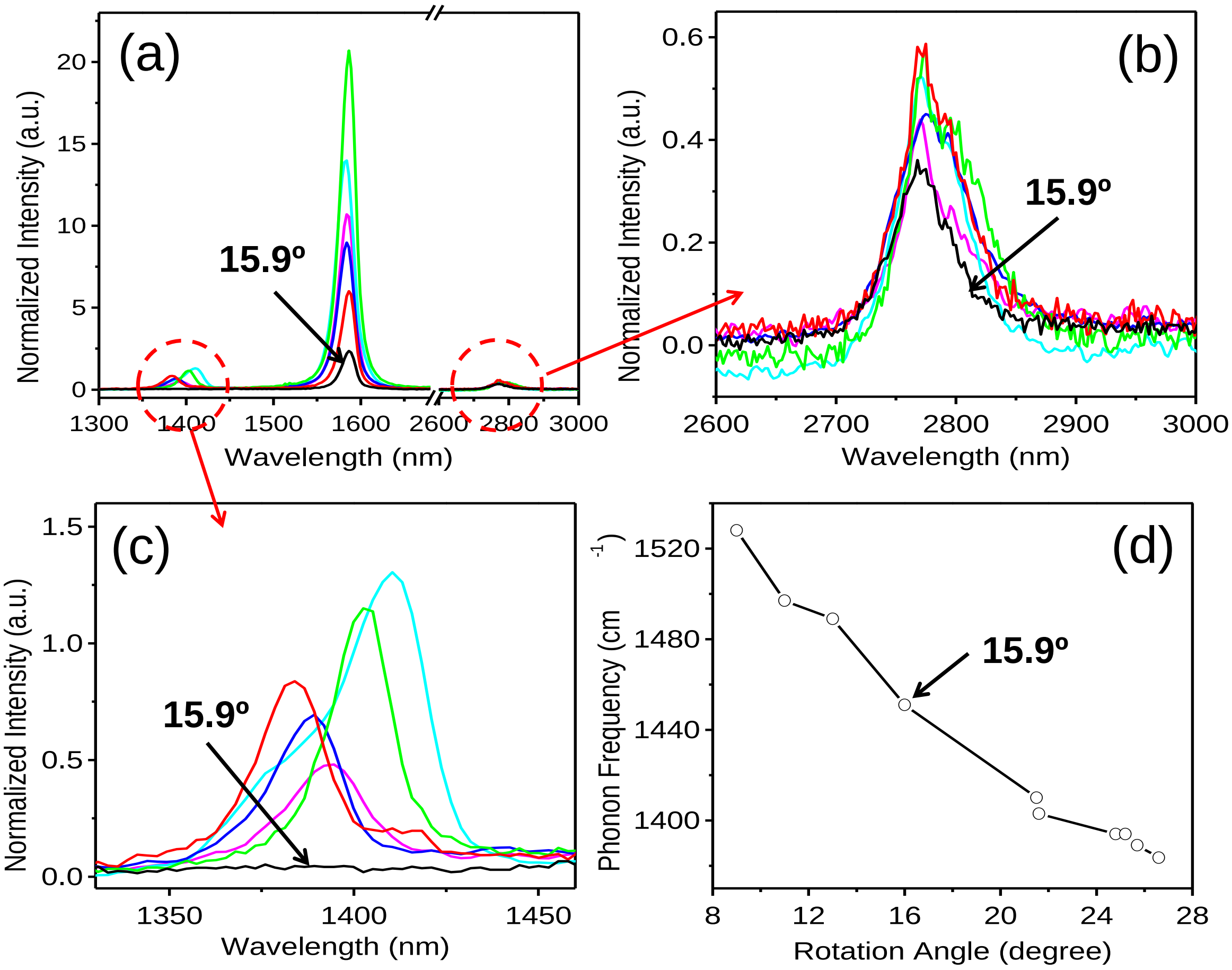

Fig. 4.

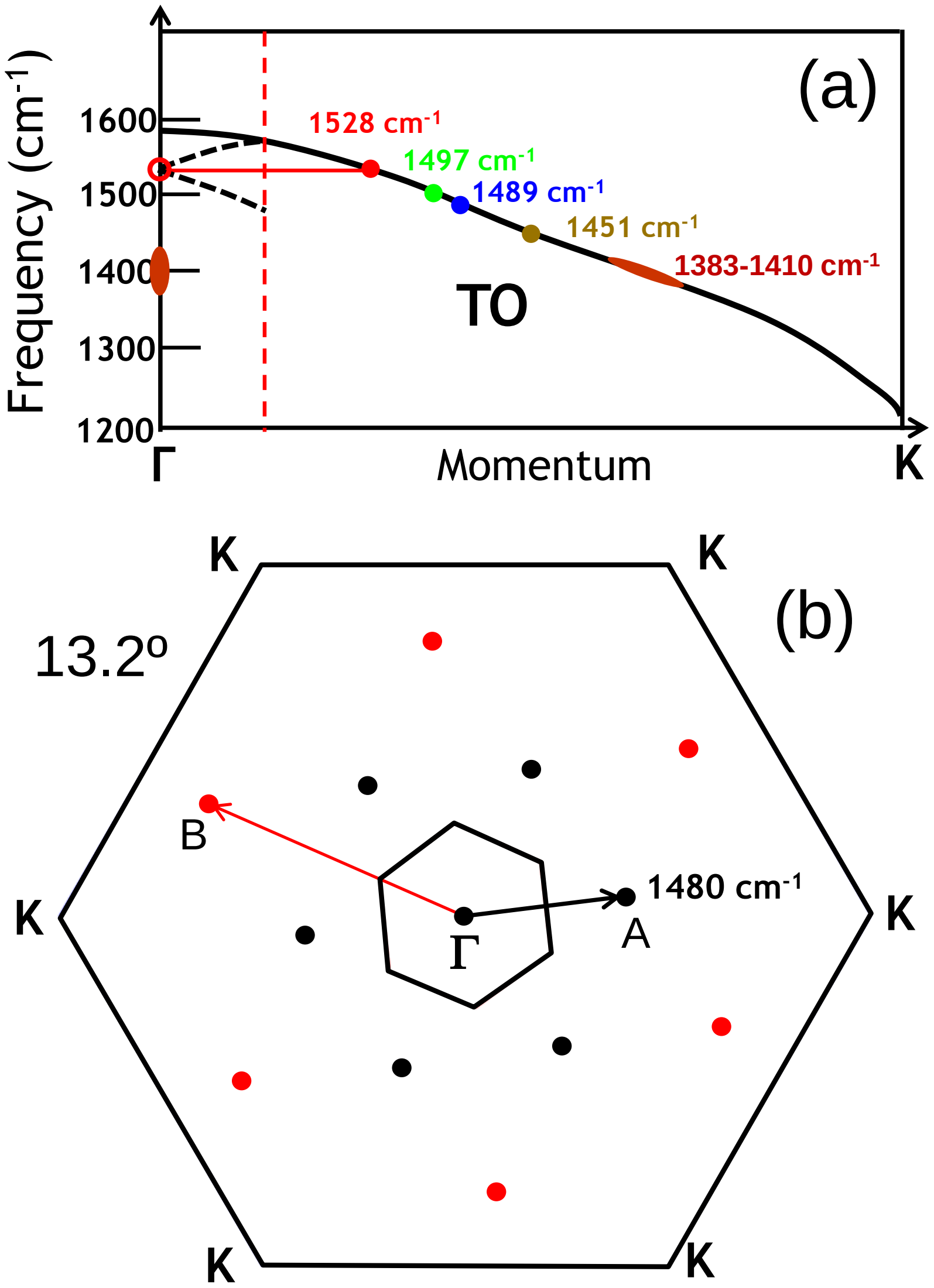

Fig. 5.

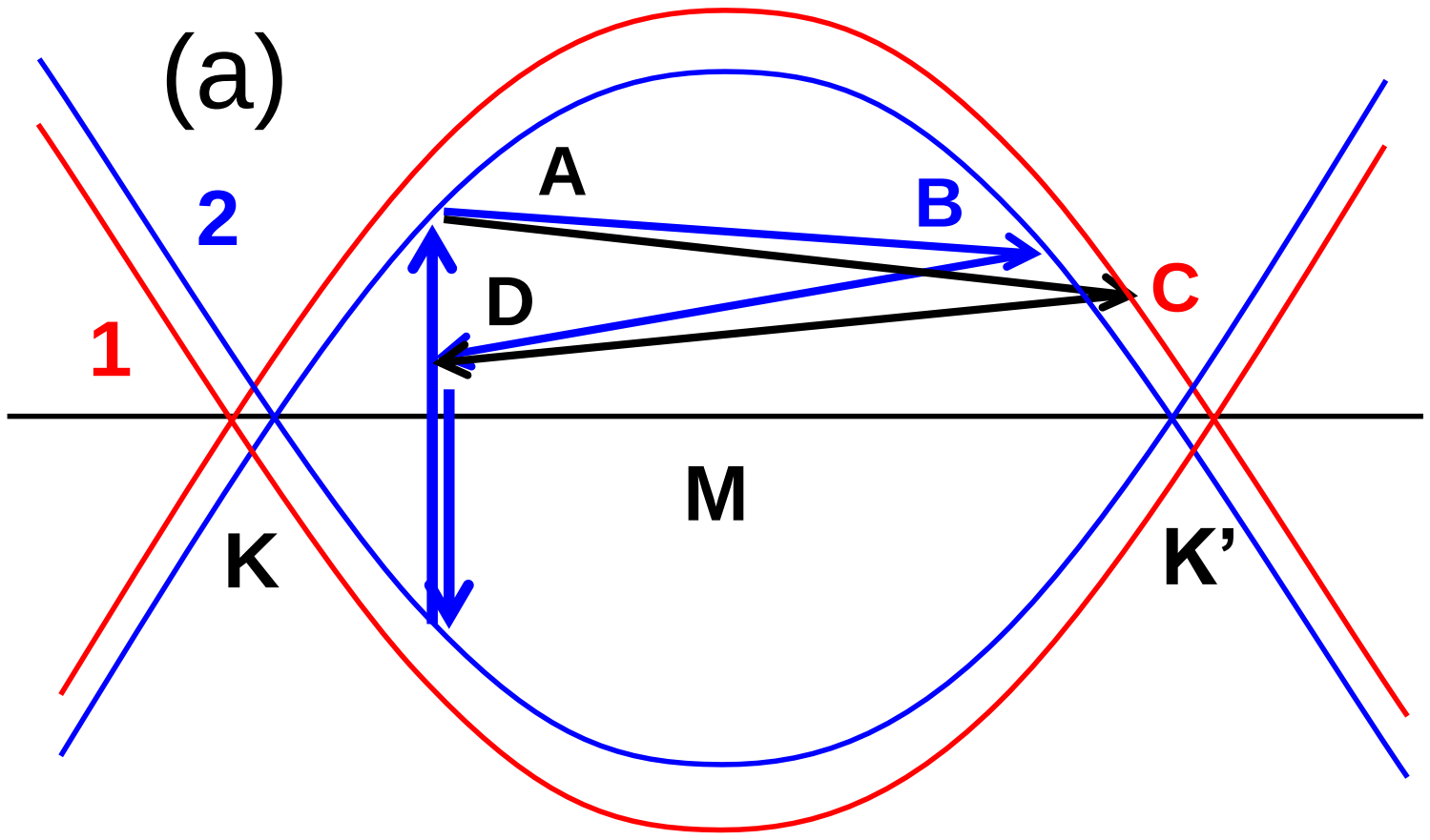

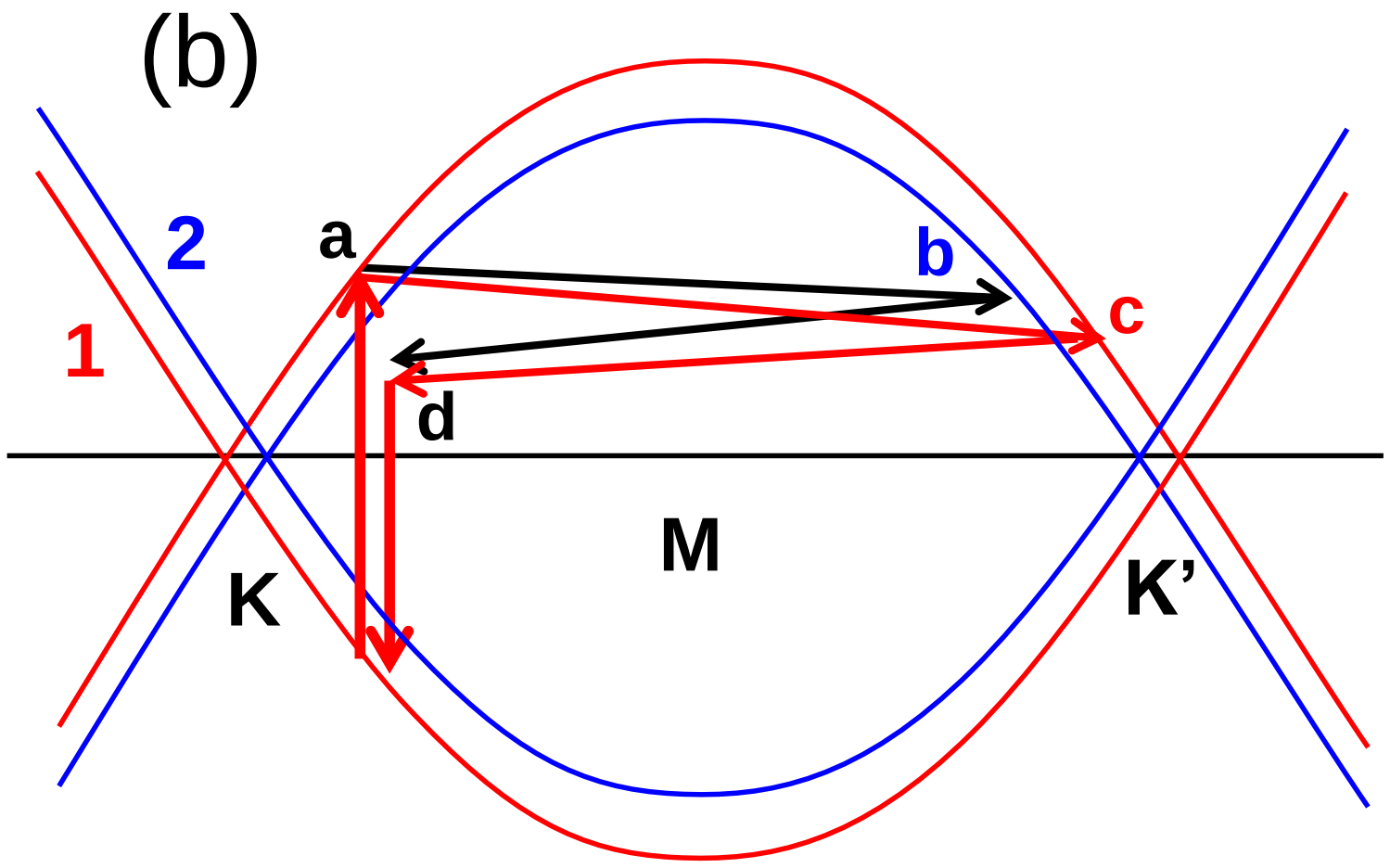